\documentclass[aps,prl, twocolumn, preprintnumbers, amsmath,amssymb,superscriptaddress, floatfix]{revtex4}
\usepackage{graphicx}
\usepackage{latexsym}
\usepackage{amsmath}
\usepackage{graphics}
\usepackage{amssymb}
\usepackage{layout}
\usepackage{verbatim}
\usepackage{bm}
\usepackage{amsfonts,epsfig}

\usepackage[varg]{txfonts}

\begin{document}
\title{Dissipation-driven quantum phase transition in superconductor-graphene systems}
\author{Roman M. Lutchyn}
\affiliation{Joint Quantum Institute and Condensed Matter Theory
Center, Department of Physics, University of Maryland, College Park,
MD 20742-4111, USA}
\author{Victor Galitski}
\affiliation{Joint Quantum Institute and Condensed Matter Theory
Center, Department of Physics, University of Maryland, College Park,
MD 20742-4111, USA}
\author{Gil Refael}
\affiliation{ Department of Physics, California Institute of
Technology, MC 114-36, Pasadena, CA 91125}
\author{S. {Das Sarma}}
\affiliation{Joint Quantum Institute and Condensed Matter Theory
Center, Department of Physics, University of Maryland, College Park,
MD 20742-4111, USA}

\date{\today}

\begin{abstract}
  We show that a system of Josephson junctions coupled via low-resistance tunneling
  contacts to graphene substrate(s) may effectively operate as a current switching device. The effect is based on the dissipation-driven
  superconductor-to-insulator quantum phase transition, which happens due to the interplay of the Josephson effect
 and Coulomb blockade. Coupling to a graphene substrate with gapless excitations  further enhances charge fluctuations
 favoring superconductivity. The effect is shown to scale
 exponentially with the Fermi energy in graphene, which can be controlled by the gate voltage.
 We develop a theory, which  quantitatively describes the quantum phase transition in a
two-dimensional Josephson junction array, but it is expected to
provide a reliable qualitative description for one-dimensional
systems as well. We argue that the local effect of
dissipation-induced enhancement of superconductivity is very
robust and a similar sharp crossover should be present in finite
Josephson junction chains.
\end{abstract}

\maketitle

 Artificially fabricated Josephson-junction arrays (JJA) offer a
unique playground for studying quantum phase transitions
(QPT)~\cite{Goldman1998}. The  transitions in JJA occur due to the
competition between the Josephson coupling, $E_J$, which favors a
globally ordered state, and the charging energy, $E_c$, which leads
to Coulomb blockade of Cooper-pair tunneling and enhances quantum
fluctuations of the superconducting (SC) phase. At zero temperature,
the QPT from a globally superconducting to an insulating
phase~\cite{Goldman1986, Mooji, Beloborodov,
Takahide_PRL00,Rimberg_PRL97} occurs, roughly, when the Josephson
energy $E_J$ becomes smaller than the charging energy. Another key
factor in determining the ground state is dissipation, which is
present whenever the SC system is connected to a reservoir of
gapless single-particle excitations~\cite{Schoen}. The main effect
of the dissipation in JJA is a suppression of quantum phase
fluctuations. Taking into account the phase-charge uncertainty
relation, the dissipation enhances fluctuations of the charge and,
hence, stabilizes the SC phase\cite{Chakravarty}. This type of
dissipative QPT has been considered previously by Feigelman and
Larkin~\cite{Feigel'man} in the model of a regular 2D
proximity-coupled JJA and by Galitski and Larkin~\cite{Galitski2001}
in a disorder-induced random Josephson network. In both cases, it
was found that the effect of dissipation on the transition point is
{\em exponential}, i.e., the critical Josephson coupling scales
exponentially with the Andreev conductance. The Andreev conductance
and hence the degree to which the stabilizing effect of dissipation
is important obviously depend on the density of states of gapless
excitations. Thus, by controlling the latter, one can tune
transitions between a global superconductor and an insulator. This
observation provides strong motivation for studying superconductors coupled
to a graphene substrate, where the density of carriers can be easily
tuned by gate voltage from essentially zero at the Dirac point (no
``Ohmic'' dissipation) to very large values in the electron-doped
graphene (strong dissipation).

\begin{figure}
\centering
\includegraphics[width=0.9\linewidth]{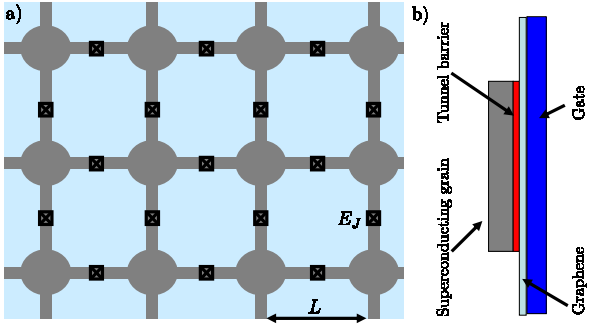}
\caption{(Color online) a) A Josephson-junction array  on top of a
graphene sheet is shown. The grains are coupled by the Josephson
junctions with the coupling strength $E_{J}$. The distance between
the grains $L$ is assumed to be large compared to the
SC coherence length $\xi$. Thus, the coherent transport
through the graphene is neglected here. Alternatively, the SC grains
in the array can be coupled to different graphene sheets. Then,
there is no coherent transport through the graphene due to physical
separation between the sheets. b) The superconductor/graphene
interface. The graphene and superconductor are connected through the
tunnel barrier. The chemical potential in graphene can be tuned with
the gate voltage $V_g$.}\label{array}
\end{figure}

In this Letter, we propose to study the superconductor-insulator
phase transition (SIT) in a Josephson-junction array in a tunneling
contact with a graphene layer (or layers) (see Fig.~\ref{array}),
which acts as a source of gapless quasiparticles. The graphene
substrate provides a unique possibility to control the dissipation
strength via the gate voltage and thereby tune the
dissipation-driven QPT.  Hence, the system may be used as a current
switching device. While the physics of the underlying effect is
intuitively quite clear, the formal description of the transition
developed in this Letter is technically non-trivial: First, we use
the tunneling Hamiltonian formalism and elements of random
 matrix theory to derive the effective phase fluctuation action of a small SC grain
 coupled to a graphene substrate. The dissipation kernel, $K(\tau)$, shows a
 crossover from the Ohmic dissipation behavior $K(\tau) \propto
 \tau^{-2}$ in the electron-doped graphene to extremely weak dissipation
$K(\tau) \propto \tau^{-4}$ at the Dirac point. Second, we develop a
mean-field theory of the SIT and show that the quantum critical
point is determined by the single-grain phase correlator, which is
to be calculated using the dissipative effective action. To
calculate the phase correlator we use the two-loop renormalization
group (RG) results from a related spin model and determine a
critical voltage, $V_c$, at which the transition occurs: For $V>
V_c$, the system is a superconductor; for $V<V_c$, the system is an
insulator.

Our theoretical model is an array of SC grains connected with each
other with the Josephson junctions and connected via tunnel contacts
to a graphene substrate, see Fig.~\ref{array}. For $T \rightarrow 0$,
one can neglect massive fluctuations of the amplitude of the order
parameter $\Delta$ in the grain, and describe the dynamics of the
system in terms of the phase-only imaginary-time effective action
($\hbar=1$)
\begin{align}
\!S_C\!+\!S_J\!=\! \label{action} \sum_i \! \int d \! \tau
\frac{\dot{\varphi}_i(\tau)^2}{E_c}\! -\! \sum_{<ij>} \!\int d \! \tau
E_J\! \cos \!\left[\varphi_i(\tau)\!-\!\varphi_j(\tau)\!\right]\!,
\end{align}
where $\varphi_i$ is the phase of the order parameter on the
$i$-th grain. Here, for simplicity, we assume that the Josephson
and charging energy are the same for all grains. However, this
assumption is not essential for our results.

We now consider the effect of graphene gapless excitations on
the phase coherence of a single SC grain. We study here the
situation when the SC grain of radius $R$ lies on top of the
graphene sheet, see Fig.~\ref{array}. In this planar geometry the
tunnel junction does not break internal symmetries of graphene, and
thus does not modify the spectrum of the excitations. In the limit
of low transparency tunnel barrier, the transport between
superconductor and graphene can be described by the tunneling
Hamiltonian
\begin{eqnarray}\label{tunneling}
H_T=t
\sqrt{d_z}\sum_{\sigma}\int_{A}{d^2\bm{r}}\left[\Psi^{\dag}_{\sigma}(\bm{r})\Psi^{(g)}_{\sigma}(\bm{r})\!+\!{\rm{H.c.}}\right],
\end{eqnarray}
where $\Psi^{(g)}_{\sigma}(\bm{r})$ and $\Psi_{\sigma}(\bm{r})$ are
the electron operators in graphene and superconductor, respectively.
Here $d_z$ is the thickness of the grain, and $t$ and $A$ are the
tunneling matrix element and the area of the junction, respectively.

\begin{figure}
\centering
\includegraphics[width=0.8\linewidth]{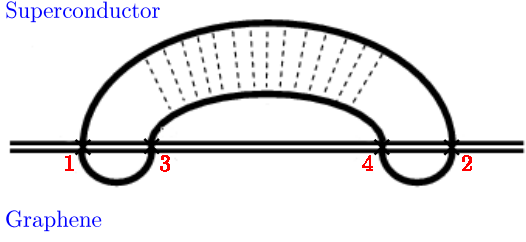}
\caption{ The diagrams describing the correlated two-electron
tunneling process (Andreev process) from/to the SC grain are
shown.}\label{diagrams}
\end{figure}

Assuming that the SC gap energy is sufficiently large, the main contribution to the subgap transport
originates from Andreev processes, which involve correlated
tunneling of two electrons from/to the graphene. In the fourth order perturbation theory in tunneling $t$ (see
Fig.~\ref{diagrams}), the contribution of the Andreev processes to
the dynamics of the phase is given by the following effective
action
\begin{align}
S_A\!&=\!24 t^4 d_z^2 \! \!\int_{A}\prod_{i=1..4} d \bm x_i \\
&\mbox{Re}\left[F^*(\bm x_1,\bm x_2) F(\bm x_3,\bm x_4) G^{(g)}(\bm
x_1, \bm x_3)G^{(g)}(\bm x_2,\bm x_4)\right],\nonumber
\end{align}
with  $\bm x=\{\bm r,\tau \}$. Here $G^{(g)}(\bm x, \bm x')$ and
$F(\bm x; \bm x')$ are  imaginary-time Green's functions for
graphene and superconductor, respectively. The latter is defined as
\begin{align}
\!& F^*(\bm r_1,\bm r_2;\tau_1,\tau_2)\!=\!-e^{i[
\phi(\tau_1)\!+\!\phi(\tau_2)]/2}\sum_n \chi_n(\bm r_1)\chi_n(\bm
r_2)u_n
v_n \nonumber\\
\!&
\left[\!\Theta(\tau_1\!-\!\tau_2)e^{-(\tau_1\!-\!\tau_2)E_n}\Theta(E_n)\!-\!\Theta(\tau_2\!-\!\tau_1)e^{-(\tau_1\!-\!\tau_2)E_n}\!\Theta(\!-\!E_n)\!\right].\nonumber
\end{align}
Here $E_n=\sqrt{\varepsilon_n^2+\Delta^2}$, $u_n$ and $v_n$ are
Bogoliubov coherence factors $u^2_n/v_n^2 = ( 1 \pm \varepsilon_n / E_n)/2$,
and $\varepsilon_n$ and $\chi_n(\bm r)$ are the eigenvalues and
eigenfunctions of the single particle Hamiltonian of the
grain, which includes random potential due to impurities and
boundaries of the island.

In order to derive low-energy effective action $S_A$ due to
Andreev processes, it is convenient to separate the fast and slow
times $\kappa$ and $\tau$: $\tau_1=\tau+\kappa/2$ and
$\tau_2=\tau-\kappa/2$. Since the superconducting Green's function
decays exponentially on the time scale of order $\Delta^{-1}$, one
can neglect $\kappa$ in the dynamics of the phase, which evolves
on much longer time scales. Taking advantage of this separation of
 scales and performing the integral over the fast time scales, we
obtain the following expression for $S_A$
\begin{align}\label{SA2}
S_A \! & \approx \! 24 t^4 d_z^2 \!\int_{A} \prod_{i=1..4} d \bm
r_i d \tau d \tau' \cos[\phi(\tau)\!-\!\phi(\tau')]
\nonumber \\
& \times F_{12} F_{34} G^{(g)}_{13}(\tau-\tau') G^{(g)}_{24}(\tau'-\tau).
\end{align}
Here $F_{ij}$ is the anomalous Green's function at
zero frequency, \emph{i.e} $F_{ij}=-\sum_n \chi_n(\bm r_i)\chi_n(\bm
r_j)\Delta/E_n^2$.

It is well-known that Andreev transport is sensitive to
disorder~\cite{Hekking1994}. Therefore, in order to calculate the
effective action, one has to take into account spatial
correlations~\cite{Hekking1994, Bruder1994, Lutchyn2007}, and
average the action over the random realization of the wavefunctions
in the SC grain resulting from the scattering of electrons from the
grain boundaries and impurities. We perform this averaging using
exact eigenstates technique assuming that the grain is sufficiently
small. Our approach accounts for the enhancement
of the tunneling rate due to the coherent back-scattering of
electrons to the tunnel junction. The correlation function $\langle
F_{12}F_{34}\rangle$ in the grain consists of reducible and
irreducible parts:
\begin{equation}
\langle F_{12}F_{34}\rangle\approx\langle F_{12}\rangle \langle
F_{34}\rangle+\langle F_{12}F_{34}\rangle_{ir}.
\end{equation}
The reducible part can be easily calculated
\begin{equation} \label{reducible}
 \langle F_{12}\rangle=\frac{\pi}{2} \nu_F f_{12} \mbox{\, with\, }
f_{12}=\frac{\sin(k_F|\bm r_1 - \bm r_2|)}{k_F|\bm r_1 - \bm
r_2|}e^{-\frac{|\bm r_1 - \bm r_2|}{2l}},
\end{equation}
where $k_F$, $l$ and $\nu_F$ are the Fermi wave vector, mean free
path, and density of states at the Fermi level in the grain,
respectively.

The irreducible correlation function is calculated assuming that SC
grains are small, and the Thouless energy $E_T$ is the largest
relevant energy scale in the problem $E_T \sim D/R^2 \gg \Delta,
E_c, E_J$, where $D$ is the diffusion constant, $R$ is the radius of
the grain, and $\delta$ is the level spacing in the grain. In this
limit, the electron diffusion time in the grain $\tau_D=1/E_{T}$ is
much smaller than the time the system dwells in the virtual state
with one unpaired electron $\sim 1/\Delta$. Since an electron in the
virtual state covers the entire available phase space, one can
calculate the irreducible correlation function within Random Matrix
Theory~\cite{Aleiner2002} finding that it acquires the universal
form $\langle F_{12}F_{34}\rangle_{ir}=\frac{\pi}{4}\frac{\nu_F^2
\delta}{\Delta}(f_{14}f_{23}+f_{13}f_{24})$. Combining this equation
with Eq.~(\ref{reducible}), we obtain
\begin{equation}\label{combined}
\langle F_{12}F_{34}\rangle=\frac{\pi^2}{4} \nu_F^2
f_{12}f_{34}+\frac{\pi}{4}\frac{ \delta}{\Delta}\nu_F^2
(f_{14}f_{23}+f_{13}f_{24}).
\end{equation}
Using Eqs.~(\ref{combined}) and (\ref{SA2}), and carrying out the
spatial integrals, we find
\begin{align}
\!\! S_A \! \approx \!  \frac{6\pi t^4}{\delta \Delta k_F^{4}}\!
\int \!d \tau d \tau' \!\cos[\phi(\tau)\!-\!\phi(\tau')]\!
G^{(g)}_{11}(\tau\!-\!\tau')G^{(g)}_{22}(\tau'\!-\!\tau).
\end{align}
This term originates from the integration of the irreducible
correlation function and is parametrically larger than  the
reducible term by the factor of $k_F^2 A \delta /\Delta \gg
1$~\cite{Hekking1994, Lutchyn2007}. After substituting the
expressions for the graphene Green's functions (see, e.g.,
Ref.~\cite{Gonzalez}) and assuming  that $\mu \ge 0$, we find
\begin{align}
\!\!\!S_A \! &\! \approx \! -G\!\! \int \!\! d \tau d \tau'
K(\tau\!-\!\tau')\! \cos \!\left[\phi(\tau)\!-\!\phi(\tau')\right],
\label{GSA} \mbox{\,} G\!=\!\frac{3t^4}{2\pi \delta \Delta k_F^{2}
\gamma^2},
\end{align}
where $\gamma$ is the Fermi-velocity in graphene. The kernel
$K(\tau)$ is described by the following function
\begin{align}
K(\tau)=\frac{1}{k_F^{2} \gamma^2
\tau^4}\!\left\{(\mu \tau)^2
 \!+\! 2 \left[1\!-\!\mu
|\tau|\right]e^{-\mu|\tau|}-1  \right\},
 \end{align}
which exhibits a crossover from $\tau^{-4}$ at the Dirac point to
the usual Ohmic behavior $\tau^{-2}$ at $\mu|\tau|\sim 1$. For
realistic experimental parameters the chemical potential $\mu$ is
much larger than the characteristic timescale for the dynamics of
the phase $\tau\leq E_c^{-1}$, i.e. $\mu \gg E_c$. Thus, the
asymptotic form of the effective action becomes
\begin{align}\label{action_A}
&\!\!\!S_A \! \approx \! - \eta \int d \tau d \tau' \frac{\cos
\left[\phi(\tau)\!-\!\phi(\tau')\right]}{(\tau-\tau')^2},\,\,\,\eta=G
\frac{\mu^2}{k_F^2 \gamma^2}
\end{align}
 The important difference between
action~(\ref{action_A}) and the dissipative action describing
resistively-shunted Josephson junction, is that action
(\ref{action_A}) is ``compact'' or periodic in the phase
difference, and thus correctly describes the fact that the charge
on the grain is quantized and can change by $2e$ only. From
Eq.~(\ref{action_A}), we see that graphene as a source of gapless
excitations provides the possibility to change the dissipation
strength directly by changing the chemical potential, which is
tied to the gate voltage.

\emph{Dissipation-driven QPT.} Combining Eqs.~(\ref{action}) and
(\ref{action_A}), we obtain the full action for the system:
$S=S_C+S_J+S_A$. To derive effective action describing the
transition in the JJA, we first write a partition function in path
integral representation and then use Hubbard-Stratonovich
transformation to decouple Josephson term by introducing an
auxiliary field $\psi_i(\tau)$ coupled linearly to
$e^{i\varphi_i(\tau)}$. Then, the partition function becomes
$Z=Z_0 \int \mathfrak{D} \psi \exp(-S[\psi])$, where $S[\psi]$ is
given by
\begin{align}\label{Spsi}
S[\psi]=&\int d \tau \frac{1}{2} \sum_{i,j} \psi^*_i(\tau)
w^{-1}_{ij} \psi_j(\tau)\\\nonumber &- \ln \left \langle \exp
\left[\frac{1}{2} \int d \tau \sum_i e^{i\varphi_i(\tau)}
\psi_i^*(\tau)+ c. c. \right] \right \rangle_0.
\end{align}
Here we introduced the symmetric matrix $w_{ij}$, which describes
Josephson tunneling: matrix elements of $w_{ij}$ are
equal to $E_{J}$ for nearest neighbors and zero otherwise. The
expectation value in Eq.~(\ref{Spsi}) is taken with respect to the
single-site action $S_0=S_C+S_A$. To study QPT at the mean field
level, one may perform cumulant
expansion of the second term in the action $S[\psi]$ in powers of
$\psi$ and arrive at the effective complex $\psi^4$ field
theory~\cite{Sachdev}. The phase boundary between macroscopically
superconducting and insulating state of JJA can be obtained by
setting the coefficient $r$ in front of $|\psi|^2$ to zero:
\begin{align}\label{r}
r \propto \frac{1}{z E_J}-\frac{1}{2}\int d\tau \langle
e^{i\varphi(\tau)-i\varphi(0)} \rangle_0=0.
\end{align}
Here $z$ is the coordination number of the lattice and averaging is
taken with respect to $S_0$. When calculating the correlation
function we assume that the dissipation strength $\eta$ is large.
Then, the second term in $S_0$ dominates at low frequencies and one
can neglect the influence of the charging energy term $S_C$, which
serves as the ultra-violet cut-off. Under these conditions, the
correlation function can be mapped on the long-range ferromagnetic
spin chain~\cite{Kosterlitz, Brezin, Hofstetter} first considered by
Kosterlitz~\cite{Kosterlitz}. Later, the critical behavior was
studied in Ref.~\cite{Hofstetter, Spohn}, where the asymptotic
behavior of the spin-spin correlation function was obtained using
two-loop renormalization group (RG). Adopting the
results of Refs.~\cite{Kosterlitz, Hofstetter, Spohn, Feigel'man} to our problem,
we get
\begin{align}\label{corr_function}
\langle e^{i\varphi(\tau)-i\varphi(0)} \rangle_0 \sim \left\{
  \begin{array}{ll}
  \displaystyle  \left({\tau_c \over \tau}\right)^{\frac{1}{2\pi^2 \eta}},
    & \Lambda^{-1} \ll \tau \ll \tau_c
  \\ \displaystyle
   \left(\frac{\tau_c}{\tau}\right)^2, & \tau \gg \tau_c, \end{array}  \right.
\end{align}
where $\Lambda$ is the ultraviolet cut-off, $\Lambda \sim 2\pi E_c
\eta$, and $\tau_c$ is the
correlation time, which can be calculated using the RG for $\eta$.
Since the compact dissipation term proportional to $\eta$ is not
Gaussian, it gets renormalized when integrating out high-frequency
degrees of freedom resulting in the following flow equations
\begin{equation}
\frac{d \eta}{d \ln (\Lambda
\tau)}=-\frac{1}{2\pi^2}-\frac{1}{(2\pi^2)^2\eta}, \label{RG1}
\end{equation}
where the rhs is the beginning of a Taylor expansion in $1/\eta$. By
integrating this equation between the initial value $\eta(0) \equiv
\eta $ and final value $2\pi^2 \eta(\ln[\Lambda\tau_c])\sim 1$, we
can estimate the correlation time as
\begin{equation}\label{tau_c}
\tau_c\sim (4 \pi^3 E_c \eta^2)^{-1}\exp\left(2\pi^2\eta\right).
\end{equation}
It is assumed here that $2\pi^2 \eta \gg 1$.
\begin{figure}
\centering
\includegraphics[width=0.9\linewidth]{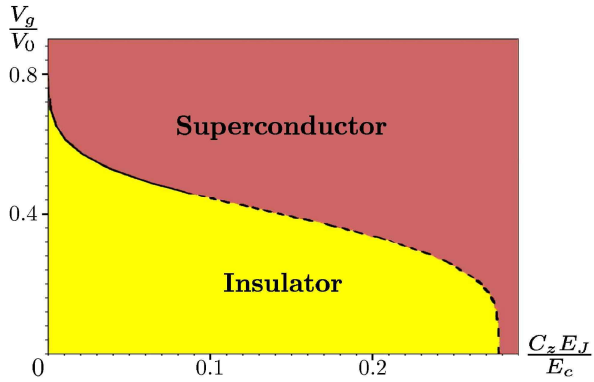}
\caption{(Color online). Phase diagram for the SIT is shown. By
tuning the applied gate voltage $V_g$, one can change the density of
states of gapless excitations in graphene, and thus drive the phase
transition between globally superconducting and insulating states of
JJA. The solid (black) line represents the region of the validity of
RG.}\label{Lambert}
\end{figure}
Using Eqs.~(\ref{corr_function}) and~(\ref{tau_c}) and carrying out
the time integral, we find expression defining the phase boundary $
C_z E_J = 4\pi^3 E_c \eta_c^2 \exp\left(-2\pi^2\eta_c\right)$, where
$C_z$ is a numerical constant depending on the coordination number
$z$. The rhs of the above equation
corresponds to the effective charging energy of the grain $E_c^*$
renormalized by the dissipation~\cite{Zaikin, Falci, Lukyanov}. The
critical dissipation strength, $\eta_c$, and the corresponding
critical voltage, $V_c$, at which the QPT occurs can be expressed
via Lambert-W function by inverting the transcendent algebraic
equation above:
\begin{align}\label{mu_c}
\frac{V_c}{V_0}=\frac{1}{\pi}\sqrt{-W_{-1}\left(- \sqrt{\pi C_z E_J
\over 4 E_c}\, \right)}.
\end{align}
Here $V_0={\gamma k_F \over e \sqrt{G}}$ with $k_F$, $\gamma$ and
$G$ being the Fermi-momentum in the SC grains, the Fermi-velocity in
graphene, and the dimensionless constant defining the transparency
of the tunnel barrier (see Eq.~(\ref{GSA})), respectively. The phase
boundary between globally superconducting and insulating phases of
JJA is shown in Fig.~\ref{Lambert}. For the theory based on the RG
procedure to be formally valid, we need $\ln (E_c/E_J) \gg 1$. This,
however, is a mathematical rather than physical constraint and for
all practical purposes (i.e., experiment) what is important is the
existence of the transition itself; the applicability of RG methods
and the exact location of the non-universal ``critical voltage'' are
not essential. A sharp transition should certainly exist if $E_c \gg
E_J$ and perhaps even for $E_c \geq E_J$. In the opposite limit the
system is already superconducting without any substrate and there is
no dissipation-driven effect. Since the technology for making SC
grains with required $E_c/E_J$ is well-developed, the conditions for
the observation of the QPT are certainly experimentally feasible.

We also emphasize here that since the dissipation-driven transition
is intrinsically local, it is expected to survive in one-dimensional
chains (where the mean-field theory breaks down and the QPT is of
Kosterlitz-Thouless type, see, e.g., Ref.~\cite{Meyer}). Again, the
exact location of the transition point and the critical behavior
would be different, but the effect itself should be present.
Moreover, the same local argument suggests that a sharp
voltage-induced crossover (cf. with the model of a shunt resistor
coupled locally to the SC grain~\cite{Refael, Tewari}) in the
IV-curves should be present even in finite chains proximity coupled
to graphene, similar to those that are already being experimentally
investigated~\cite{Delft, Andrei}. We propose that experiments be carried out in the SC-graphene system to directly confirm our quantum phase transition and current switching predictions.

{\it Acknowledgments.} We thank M.~Feigel'man, E.~Hwang, J.~Lau and
S.~Tewari for stimulating discussions. VG acknowledges the
hospitality of Boston University visitors program. This work has been supported (RML and SDS) by US-ONR and LPS-NSA-CMTC.

\end{document}